\documentclass[twoside,twocolumn,9pt]{article}
\usepackage{extsizes}
\usepackage[super,sort&compress,comma]{natbib} 
\usepackage[version=3]{mhchem}
\usepackage[left=1.5cm, right=1.5cm, top=1.785cm, bottom=2.0cm]{geometry}
\usepackage{balance}
\usepackage{mathptmx}
\usepackage{sectsty}
\usepackage{graphicx} 
\usepackage{lastpage}
\usepackage[format=plain,justification=justified,singlelinecheck=false,font={stretch=1.125,small,sf},labelfont=bf,labelsep=space]{caption}
\usepackage{float}
\usepackage{fancyhdr}
\usepackage{fnpos}
\usepackage[english]{babel}
\addto{\captionsenglish}{%
  
}
\usepackage{array}
\usepackage{droidsans}
\usepackage{charter}
\usepackage[T1]{fontenc}
\usepackage[usenames,dvipsnames]{xcolor}
\usepackage{setspace}
\usepackage[compact]{titlesec}
\usepackage{hyperref}
\usepackage{upgreek}
\usepackage{epstopdf}%This line makes .eps figures into .pdf - please comment out if not required.

\definecolor{cream}{RGB}{222,217,201}

\begin{document}

\pagestyle{fancy}
\thispagestyle{plain}
\fancypagestyle{plain}{
%%%HEADER%%%
\renewcommand{\headrulewidth}{0pt}
}
\newcommand{\microns}{\, \rm \upmu m}
\newcommand{\equi}{\mathrm{eq}}
\newcommand{\rp}{\mathrm{R}}
\newcommand{\lp}{\mathrm{L}}
\newcommand{\lr}{\mathrm{L,R}}

%%%END OF HEADER%%%

%%%PAGE SETUP - Please do not change any commands within this section%%%
\makeFNbottom
\makeatletter
\renewcommand\LARGE{\@setfontsize\LARGE{15pt}{17}}
\renewcommand\Large{\@setfontsize\Large{12pt}{14}}
\renewcommand\large{\@setfontsize\large{10pt}{12}}
\renewcommand\footnotesize{\@setfontsize\footnotesize{7pt}{10}}
\makeatother

\renewcommand{\thefootnote}{\fnsymbol{footnote}}
\renewcommand\footnoterule{\vspace*{1pt}% 
\color{cream}\hrule width 3.5in height 0.4pt \color{black}\vspace*{5pt}} 
\setcounter{secnumdepth}{5}

\makeatletter 
\renewcommand\@biblabel[1]{#1}            
\renewcommand\@makefntext[1]% 
{\noindent\makebox[0pt][r]{\@thefnmark\,}#1}
\makeatother 
\renewcommand{\figurename}{\small{Fig.}~}
\sectionfont{\sffamily\Large}
\subsectionfont{\normalsize}
\subsubsectionfont{\bf}
\setstretch{1.125} %In particular, please do not alter this line.
\setlength{\skip\footins}{0.8cm}
\setlength{\footnotesep}{0.25cm}
\setlength{\jot}{10pt}
\titlespacing*{\section}{0pt}{4pt}{4pt}
\titlespacing*{\subsection}{0pt}{15pt}{1pt}
%%%END OF PAGE SETUP%%%

%%%FOOTER%%%
\fancyfoot{}
\fancyfoot[LO,RE]{\vspace{-7.1pt}\includegraphics[height=9pt]{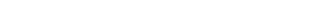}}
\fancyfoot[CO]{\vspace{-7.1pt}\hspace{13.2cm}\includegraphics{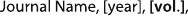}}
\fancyfoot[CE]{\vspace{-7.2pt}\hspace{-14.2cm}\includegraphics{head_foot/RF}}
\fancyfoot[RO]{\footnotesize{\sffamily{1--\pageref{LastPage} ~\textbar  \hspace{2pt}\thepage}}}
\fancyfoot[LE]{\footnotesize{\sffamily{\thepage~\textbar\hspace{3.45cm} 1--\pageref{LastPage}}}}
\fancyhead{}
\renewcommand{\headrulewidth}{0pt} 
\renewcommand{\footrulewidth}{0pt}
\setlength{\arrayrulewidth}{1pt}
\setlength{\columnsep}{6.5mm}
\setlength\bibsep{1pt}
%%%END OF FOOTER%%%

%%%FIGURE SETUP - please do not change any commands within this section%%%
\makeatletter 
\newlength{\figrulesep} 
\setlength{\figrulesep}{0.5\textfloatsep} 

\newcommand{\topfigrule}{\vspace*{-1pt}% 
\noindent{\color{cream}\rule[-\figrulesep]{\columnwidth}{1.5pt}} }

\newcommand{\botfigrule}{\vspace*{-2pt}% 
\noindent{\color{cream}\rule[\figrulesep]{\columnwidth}{1.5pt}} }

\newcommand{\dblfigrule}{\vspace*{-1pt}% 
\noindent{\color{cream}\rule[-\figrulesep]{\textwidth}{1.5pt}} }

\makeatother
%%%END OF FIGURE SETUP%%%

%%%TITLE, AUTHORS AND ABSTRACT%%%
\twocolumn[
  \begin{@twocolumnfalse}
{\includegraphics[height=30pt]{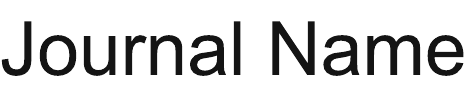}\hfill\raisebox{0pt}[0pt][0pt]{\includegraphics[height=55pt]{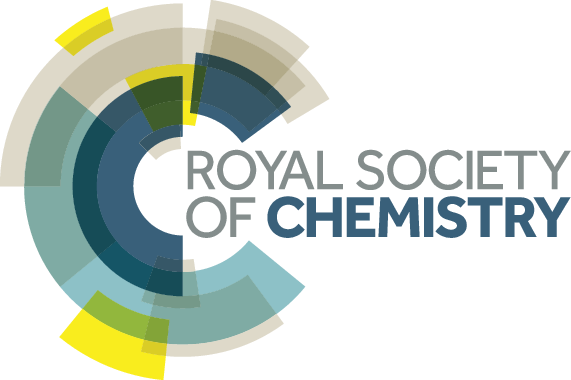}}\\[1ex]
\includegraphics[width=18.5cm]{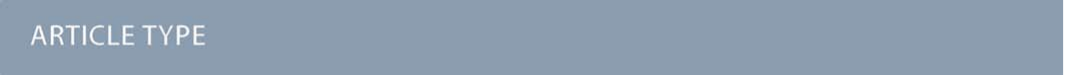}}\par
\vspace{1em}
\sffamily
\begin{tabular}{m{4.5cm} p{13.5cm} }

\includegraphics{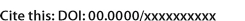} & 
\noindent\LARGE{\textbf{Probing the chirality of a single microsphere trapped by a focused vortex beam through their orbital period}}\\ 
\vspace{0.3cm} & \vspace{0.3cm} \\

 & \noindent\large{Kain\~a Diniz,\textit{$^{a,b}$}$^{\dag}$ Tanja Schoger,\textit{$^{c \dag}$} Arthur L. Fonseca,\textit{$^{a,b}$}, 
 Rafael S. Dutra,\textit{$^{d}$}
 Diney S. Ether Jr,\textit{$^{a,b}$} Gert-Ludwig Ingold,\textit{$^{c}$} 
 Felipe A. Pinheiro,\textit{$^{a}$}
 Nathan B. Viana,\textit{$^{a,b}$} 
 and Paulo A. Maia Neto\textit{$^{\ast a,b}$}} \\

\includegraphics{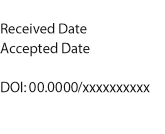} & \noindent\normalsize{When microspheres are illuminated by tightly focused vortex beams, they can be trapped in a non-equilibrium steady state where they orbit around the optical axis. 
By using the Mie-Debye theory for optical tweezers, we demonstrate that the orbital period strongly depends on the particle's chirality index. 
Taking advantage of such sensitivity, we put forth a method to experimentally characterize with high precision the chiroptical response of individual optically trapped particles. The 
method allows for an enhanced precision at least one order of magnitude larger than that of similar existing enantioselective approaches. It is particularly suited to probe the chiroptical response of individual particles, for which light-chiral matter
interactions are typically weak.} \\

\end{tabular}

 \end{@twocolumnfalse} \vspace{0.6cm}

]
%%%END OF TITLE, AUTHORS AND ABSTRACT%%%

%%%FONT SETUP - please do not change any commands within this section
\renewcommand*\rmdefault{bch}\normalfont\upshape
\rmfamily
\section*{}
\vspace{-1cm}

%%%FOOTNOTES%%%
\footnotetext{\textit{$^{\dag}$~These authors contributed equally to this work.}}

\footnotetext{\textit{$^{a}$~Instituto de F\'{\i}sica, Universidade Federal do Rio de Janeiro, Caixa Postal 68528, Rio de Janeiro,  Rio de Janeiro, 21941-972, Brazil; Email: dinizkaina@gmail.com}}

\footnotetext{\textit{$^{b}$~CENABIO - Centro Nacional de Biologia Estrutural e Bioimagem, Universidade Federal do Rio de Janeiro,
Rio de Janeiro, Rio de Janeiro, 21941-902, Brazil}}

\footnotetext{\textit{$^{c}$~Institut für Physik, Universität Augsburg, 86135 Augsburg, Germany}}

\footnotetext{\textit{$^{d}$~LISComp-IFRJ, Instituto Federal de Educa\c c\~ao, Ci\^encia e Tecnologia, Rua Sebast\~ao de Lacerda, Paracambi, Rio de Janeiro, 26600-000, Brasil}}

%Please use \dag to cite the ESI in the main text of the article.
%If you article does not have ESI please remove the the \dag symbol from the title and the footnotetext below.
%\footnotetext{\dag~Electronic Supplementary Information (ESI) available: [details of any supplementary information available should be included here]. See DOI: 00.0000/00000000.}
%additional addresses can be cited as above using the lower-case letters, c, d, e... If all authors are from the same address, no letter is required

%\footnotetext{\ddag~Additional footnotes to the title and authors can be included \textit{e.g.}\ `Present address:' or `These authors contributed equally to this work' as above using the symbols: \ddag, \textsection, and \P. Please place the appropriate symbol next to the author's name and include a \texttt{\textbackslash footnotetext} entry in the correct place in the list.}

%%%MAIN TEXT%%%%
\section{Introduction}
\label{sec:introduction}

Chiral discrimination plays a crucial role in many areas of science such as Chemistry, Molecular Biology and Pharmaceutics (see \textit{e.g.} Ref.~\citenum{maier2001} for review).
Over the years, various methods to separate molecules and particles based on their chiral properties were developed. 
There exist chemical processes to separate enantiomers from each other (see  \textit{e.g.} Refs.~\citenum{Gerald2001, Scriba2002} for reviews), which, however, have the disadvantage that they are usually developed for specific chiral particles and tend to be invasive. In addition, they usually probe only the average chiral response of an ensemble of chiral particles or molecules, rather than that of individual particles, for which such response is typically small.~\cite{kumar2020recent,solomon2020nanophotonic} To circumvent this limitation, plasmonic nanostructures have been used in enantioselective schemes due to their ability to enhance chiroptical properties based on localized surface plasmon resonance.~\cite{kim2022enantioselective,liu2020enantiomeric}
Recently, all-optical chiral discrimination methods have received significant attention due to their potential as noninvasive alternatives\cite{Genet_2022,ben2013chirality}, and because they are particularly suited to characterize the chiral response of single, isolated chiral nanoparticles~\cite{zhao2016enantioselective,ho2017enhancing,schnoering2018three,shi2020chirality,lee2024unraveling,ali2023enantioselective,ali2021enantioselection,AliJosaB,Ali_Nanoscale}.
These methods are possible because chiral particles respond differently to left- and right-circularly polarized light.\cite{Canaguier-Durand_2013, Genet_2022} This has been exploited, for instance, in the context of optical tweezers,~\cite{ashkin1986,Gennerich2016} with several methods being introduced in recent years to trap and characterize single chiral particles using tightly focused beams.~\cite{AliJosaB, Ali_Nanoscale,  Zhang_2022, Yamanishi2022}
The proposal in Refs.~\citenum{AliJosaB, Ali_Nanoscale} is based on an optical torque which the particle experiences when displaced from its equilibrium position on the optical axis by an external force. 
Due to focusing, the spin angular momentum associated with polarization can be exchanged with the trapped particle as orbital angular momentum\cite{Zhao2007, Nieminen2008}, generating a torque that is sensitive to the chirality of the particle.

In addition to spin, light can also carry intrinsic orbital angular momentum, which is associated with the field's phase distribution in space.\cite{Allen1992} Paraxial beams that carry this type of angular momentum are called vortex beams.
An important class of vortex beams are the Laguerre-Gaussian modes, usually denoted by LG$_{p\ell}$, where $p$ is a positive integer which determines the number of radial nodes, and $\ell$ is an integer called topological charge. 
In addition to spin angular momentum associated with polarization, such modes carry an orbital angular momentum of $\ell \hbar$ per photon related to their helix-shaped wavefront, with the sign of $\ell$ determining the direction of the twist of the helices.
Upon interaction with such paraxial fields, a chiral dipole cannot discriminate between different topological charges. \cite{Andrews2004,Araoka2005} 
An experiment with tightly focused vortex beams showed also no response of chiral molecules on beams with different topological charges. \cite{Loeffler2011}
However, more recent studies revealed that chiral materials indeed respond in a discriminatory way to the handedness and magnitude of light's orbital angular momentum because of quadrupole contributions.\cite{Brullot2016, Kerber2018} 
If the field becomes strongly focused, the spin and orbital degrees of freedom become coupled~\cite{Bliokh2015}, and a chiral particle’s response will be different for different topological charges.\cite{Li_2021, Zhang_2022}

In the context of optical trapping, focusing of vortex beams with $\ell \neq 0$ leads to a ring-shaped focal spot. 
If a particle is small compared to the diameter of the ring, it can be trapped in a non-equilibrium steady state where it orbits around the optical axis.\cite{Curtis2003,Simpson2010,Chen2013}
For brevity, we refer to this type of state as the ring-trapping regime in the following.
Li \textit{et al.} \cite{Li_2021} found that, for a particle confined to the focal plane, both the radius of the orbit and the optical torque that drives the particle depend on its chirality. Also, it has recently been shown that optical tweezers with vortex beams with $\ell \neq 0$ exert an enhanced torque upon trapped objects, and that this effect can be used to characterize material properties of microspheres. \cite{diniz2023}
Here we propose to use the period of particles in the ring regime as a probe for their chirality. Beyond the usual discussion about enantioselectivity, we present a proposal to quantify microsphere's chirality while estimating the resolution that could be achieved.
Additionally, by calculating the radius of the orbit and its location along the axis from the conditions of 
vanishing axial and radial force components, we provide a more realistic model when compared to the ones which consider the azimuthal force only in the focal plane. 
We also demonstrate that, in our scenario, analyzing the period yields a higher chiral resolution than doing so with just the orbital radius. This result is particularly suited for enantioselection of individual particles, where chiroptical response is typically small, and for this reason our proposal singles out with respect to other existing enantioselective methods for single chiral particles.

\section{Mie-Debye theory for chiral nanospheres trapped by a vortex beam}

To describe the response of a chiral particle to an electromagnetic field, we use the following set of constitutive equations\cite{Shivola91}
\begin{equation}
  \begin{pmatrix}
  \mathbf{D} \\ \mathbf{B} 
  \end{pmatrix} = 
  \begin{pmatrix}
      \epsilon_0 \epsilon  & i \kappa/c \\
      - i\kappa/c & \mu_0 \mu
  \end{pmatrix}
  \begin{pmatrix}
  \mathbf{E} \\ \mathbf{H}
  \end{pmatrix}\,,
  \label{eq:constitutive_equation}
\end{equation}
where $\epsilon$ and $\mu$ are the relative permittivity and permeability,
$c = 1/\sqrt{\epsilon_0 \mu_0}$ is the vacuum speed of light, and $\kappa$ is a pseudo-scalar known as the chirality parameter. 
Although these equations assume a homogeneous and isotropic response, particles whose chirality arises from their geometry can also be considered in terms of an effective chirality parameter.\cite{Mun2020}
Notice that $\kappa$ accounts for an electro-to-magnetic and magneto-to-electric coupling.

To describe the trapping of a chiral spherical particle of radius $R$ by a tightly focused vortex beam, we have developed a version of the Mie-Debye theory for optical tweezers with Laguerre-Gaussian modes \cite{Fonseca2023} that includes chiral scatterers \cite{ali2020sphere, AliJosaB}. 
The field before focusing is assumed to be a circularly polarized ($\sigma = \pm  1$) Laguerre-Gaussian beam  LG$_{0\ell}$ with one intensity node and topological charge $\ell$.
The angular spectrum representation of the electric field resulting from the focusing of such a beam by an objective is given by \cite{monteiro2009, diniz2023}
\begin{equation}
\begin{aligned}
\mathbf{E}^{(\sigma, \ell)}(\mathbf{r}) &= -\frac{ikf E_0 e^{-ikf}}{2\pi} \left(\frac{\sqrt{2}f}{w_0} \right)^{|\ell|}
	\int_0^{2\pi} d\varphi \, e^{i\ell \varphi} \\
	& \quad \times
	\int_0^{\theta_0} d\theta\, \sin\theta \sqrt{\cos\theta} 
	\sin^{|\ell|}(\theta) 	
	e^{- (f\sin(\theta)/w_0)^2} \\
	& \hspace{4em} \times
	e^{i\mathbf{k} \cdot \mathbf{r}}
	\hat{\epsilon}_\sigma (\theta, \varphi)\,.
\end{aligned}
\label{eq:E_inc}
\end{equation}
The integral covers the direction of all wave vectors $\mathbf{k} = \mathbf{k}(k, \theta, \varphi)$ within the medium of refractive index $n_\text{w}$ surrounding the sphere, up to a maximal angle defined by $\sin(\theta_0) = \mathrm{NA}/n_\text{w}$, where NA is the numerical aperture of the objective that performs the focusing. 
The wave number $k= 2\pi n_\text{w}/\lambda_0$ is defined in terms of the vacuum wavelength $\lambda_0$ of the beam. $E_0$ denotes the field amplitude, while $f$ defines the focal length and $w_0$ the beam waist at the entrance of the objective. 
The polarization unit vector is given by $\hat{\epsilon}_\sigma(\theta, \varphi) = e^{i\sigma \varphi}(\hat{\theta}
 + i\sigma \hat{\varphi})/\sqrt{2}$ where $\hat{\theta}$ and $\hat{\varphi}$ refer to the unit vectors in spherical coordinates.
We obtain the scattered field by applying Mie theory.
The optical force $\mathbf{F}$ exerted by the total field can be calculated by integrating the time-averaged Maxwell's stress tensor over a closed surface around the spherical scatterer.
In the context of the Mie-Debye theory, rather than working directly with the force, it is convenient to define the dimensionless quantity $\mathbf{Q}$ called efficiency factor\cite{ashkin1986}
\begin{equation}\label{Eq:Qfactor_def}
	\mathbf{Q} = \frac{\mathbf{F}}{(n_\text{w}/c)P}\,,
\end{equation}
where $P$ is the power on the sample. 
The efficiency factor quantifies the force exerted by the field upon the particle per unit power. 
Due to the axial symmetry of the vortex beam, it is convenient to express the optical force in cylindrical coordinates $\mathbf{Q} = Q_\rho \hat{\rho} + Q_\phi \hat{\phi} + Q_z \hat{z}$. 
The component $Q_z$ defines the axial force along the propagation direction of the beam, while $Q_\rho$ and $Q_\phi$ are the transverse force components in the radial and azimuthal direction, respectively.
Furthermore, we also define the position $(\rho, z, \phi)$ of the sphere with respect to the focus in cylindrical coordinates.
The explicit force expressions for a trapped dielectric sphere can be found in Refs.~\citenum{Fonseca2023, diniz2023}. 
For a chiral sphere, the electric and magnetic Mie scattering coefficients $a_j$ and $b_j$ of multipole order $j$ have to be replaced by
\begin{equation}
a_j \rightarrow a_j + i\sigma d_j\,, \qquad
b_j \rightarrow b_j - i\sigma c_j\, .
\end{equation}
The scattering coefficients for a size parameter $x=kR$ are given by 
\begin{align}
a_j(x) &=  \Delta^{-1}_j(x) \left[V_j^\rp(x) A_j^\lp(x) + V_j^\lp(x) A_j^\rp(x)\right]\, , \\
b_j(x) &=  \Delta^{-1}_j(x) \left[W_j^\rp(x) B_j^\lp(x) + W_j^\lp(x) B_j^\rp(x) \right]\, , \\
c_j(x) &= - d_j(x) =   i\Delta^{-1}_j(x) \left[W_j^\rp(x) A_j^\lp(x) - W_j^\lp(x) A_j^\rp(x)\right]\, ,
\end{align}
where we used the following auxiliary functions
\begin{equation}
\Delta_j(x) = W_j^\lp (x) V_j^\rp (x) + W_j^\rp (x) V_j^\lp (x)\, ,
\end{equation}
\begin{align}
W_j^\lr (x) &= M\psi_j(N_\lr x) \xi'_j(x) -       \xi_j(x)\psi'_j(N_\lr x)\,, \\
V_j^\lr(x) &=   \psi_j(N_\lr x) \xi'_j(x) - M \xi_j(x)\psi'_j(N_\lr x)\, , \\
A_j^\lr(x) &= M\psi_j(N_\lr x) \psi'_j(x) -      \psi_j(x)\psi'_j(N_\lr x)\, , \\
B_j^\lr(x) &=   \psi_j(N_\lr x) \psi'_j(x) - M\psi_j(x)\psi'_j(N_\lr x)\,
\end{align} 
with the relative impedance $M= n_\text{w} \sqrt{\mu/\epsilon}$ and the relative refractive index $N_\lr = (n \pm \kappa)/n_\text{w}$, $n = \sqrt{\epsilon \mu}$ for left (L) and right (R) polarized waves. 
Note that we adapted the notation from Ref.~\citenum{bohren1974}, where similar Mie coefficients were obtained, but for different constitutive equations than the ones presented in Eq.~\eqref{eq:constitutive_equation}.
The Mie coefficients for the scattered field are expressed in terms of the Riccati-Bessel functions $\psi_j(z) = z j_j(z)$ and $\xi_j(z) = z h_j^{(1)}(z)$, where $j_j(z)$ and $h_j^{(1)}(z)$ are the spherical Bessel and Hankel functions of the first kind, respectively.
Due to the reciprocity of chiral materials, the polarization-mixing coefficients fulfill $c_j = -d_j$. 
If the chirality parameter vanishes, the coefficients reduce to the usual Mie coefficients $a_j = A_j/W_j, b_j = B_j/V_j$ and $c_j =0 = d_j$.

\section{Results and discussion}
\label{sec:method_period}

\begin{figure}
    \centering    \includegraphics[width=0.45\textwidth]{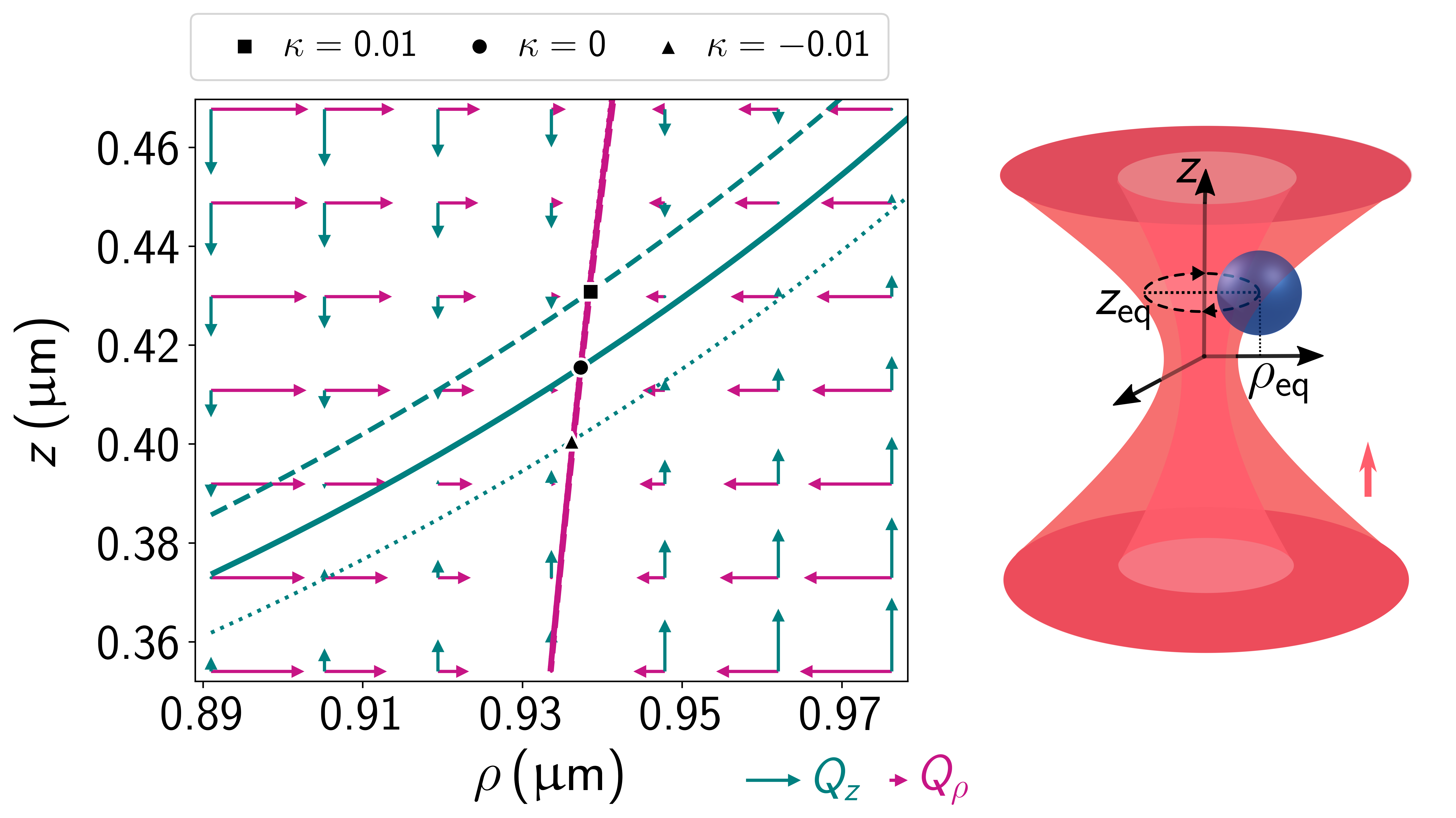}
    \caption{Left panel: Optical force field in the $\rho-z$ plane for an achiral sphere of radius $0.3\microns$ in a focused vortex beam with topological charge $\ell = 4$. The purple vectors represent the radial force component $Q_\rho$ and the teal-colored vectors denote the axial force components. The axial force components are scaled by a factor of three compared to the radial force component. 
The solid purple- and teal-colored lines represent the vanishing of the radial and axial force components, respectively. The intersection (circle symbol) defines the cylindrical coordinates of the orbit
($\rho_\equi, z_\equi$). 
For comparison, the zero-force curves for a sphere with chirality index $\kappa = 0.01$ (dashed lines) and $\kappa = -0.01$ (dotted lines) are also shown. Note that the lines for zero axial force component are too close to be distinguishable, reflecting its weak dependence with $\kappa.$ The orbit coordinates corresponding to $\kappa = 0.01$ and $\kappa = -0.01$ are indicated by the square and the triangle, respectively.
Right panel: Schematic representation of trapping of a sphere above the focal plane with the arrow indicating the propagation direction of the light beam.}
    \label{fig:vectorfield}
\end{figure}

Using the Mie-Debye theory for chiral particles outlined above, we examine the period of stably trapped objects.
A particle in a steady-state orbit around the optical axis is in equilibrium in the axial and radial directions, which means that the optical force components in those two directions must vanish, as illustrated in Fig.~\ref{fig:vectorfield}.
To find the coordinates of the circular orbit $\rho_\equi$ and $z_\equi$, we use the Mie-Debye theory to simultaneously solve the equations
\begin{equation}\label{Eq:EqZcondition}
    Q_z(\rho_\equi, z_\equi) = 0\,,
\end{equation}
\begin{equation}\label{Eq:EqRhocondition}
    Q_{\rho}(\rho_\equi, z_\equi) = 0\,.
\end{equation}
We also require that the derivatives $\partial_\rho Q_\rho$ and $\partial_z Q_z$ at
$(\rho_\equi,z_\equi)$ are negative to ensure that the orbit is stable.

As the particle is typically immersed in some fluid, it experiences a drag force proportional to its speed. \cite{Feitosa1991, Schaffer2007} 
The particle will perform a uniform circular motion whose speed $v_\phi$ is such that the drag force and the azimuthal component of the optical force cancel each other. Thus, using the definition \eqref{Eq:Qfactor_def}, we find the following relation for the orbiting speed
\begin{equation}
    v_{\phi} = \frac{n_\text{w}P}{c \gamma} Q_{\phi}(\rho_\equi, z_\equi)\, ,
\end{equation}
where $\gamma$ is the Stokes drag coefficient, \textit{i.e.}, the proportionality constant between the particle speed and the drag force.
Together with the relation $v_\phi = \rho_\equi \omega$ between the velocity and angular velocity $\omega$, the period $T=2\pi/\omega$ can be expressed as
\begin{equation}\label{Eq:period}
T = \frac{2\pi\rho_\equi\gamma}{(n_{\text{w}} P/c) Q_{\phi}(\rho_\equi, z_\equi)}\,.
\end{equation}
We characterize the liquid by a viscosity $\eta$ and account for the influence of the walls of the sample by applying the Faxén correction to the drag coefficient of a spherical particle 
\cite{Brenner1961}
\begin{equation}
    \gamma = \frac{6\pi\eta R}{1 - \frac{9}{16} \frac{R}{h} 
    + \frac{1}{8}\left(\frac{R}{h}\right)^3 
    - \frac{45}{256}\left(\frac{R}{h}\right)^4
    - \frac{1}{16}\left(\frac{R}{h}\right)^5} \, ,
\end{equation}
where $h$ is the distance of the sphere's center from the interface.

We analyze the period, as given by Eq.~\eqref{Eq:period}, and its dependence on the chirality index of a sphere for vortex beams with various topological charges.
The beam is assumed to be left-circularly ($\sigma =1$) polarized. 
For all numerical results discussed below, we assume an objective with numerical aperture $\rm NA = 1.2$ and back aperture radius $R_{\rm obj} = 2.8 \, \rm mm$, values that are typical for commercially available objectives. 
To make fair comparisons between different modes, one must ensure that they have similar filling conditions at the objective entrance port. 
Thus, for each Laguerre-Gaussian mode, except when $\ell=0$, we compute the beam waist such that the ratio between the radius of the ring of maximal intensity and the objective equals 0.8, as it is described in detail in Ref.~\citenum{Fonseca2023}. 
This implies that the beam waist is given by  $w_0(\ell) = 0.8 R_{\rm obj} \sqrt{2/|\ell|}$ for $\ell\neq 0$.
We note that this type of dynamic waist control can be performed with light modulation devices\cite{diniz2023}, the same that can generate vortex beams. 
For the Gaussian mode, we set $w_0(\ell=0)=2.2\,\mathrm{mm}$. 
The microsphere center of mass is always taken to be $h=$ $2 \microns$ above the coverslip. 
We consider a non-magnetic scatterer with a refractive index $n = \sqrt{\epsilon} = 1.57$ for a vacuum wavelength $\lambda_0 = 1064$\,nm, so as to emulate a polystyrene microsphere. \cite{zhang2020} 
The refractive index of water is $n_\text{w}=1.32$. \cite{daimon2007} 
For the calculation of the Stokes drag force, we use the viscosity of water at $20\, ^\circ$C, which is $\eta = 1.0016\, \mathrm{m Pa \cdot s}$.
The power at the sample is set to $P=10$\,mW. 
As we will see later, the choice of $P$ in our method is a matter of experimental convenience, and the theoretical resolution for $\kappa$ measurements is not directly affected by it.

\begin{figure}
    \centering
    \includegraphics[width=0.45\textwidth]{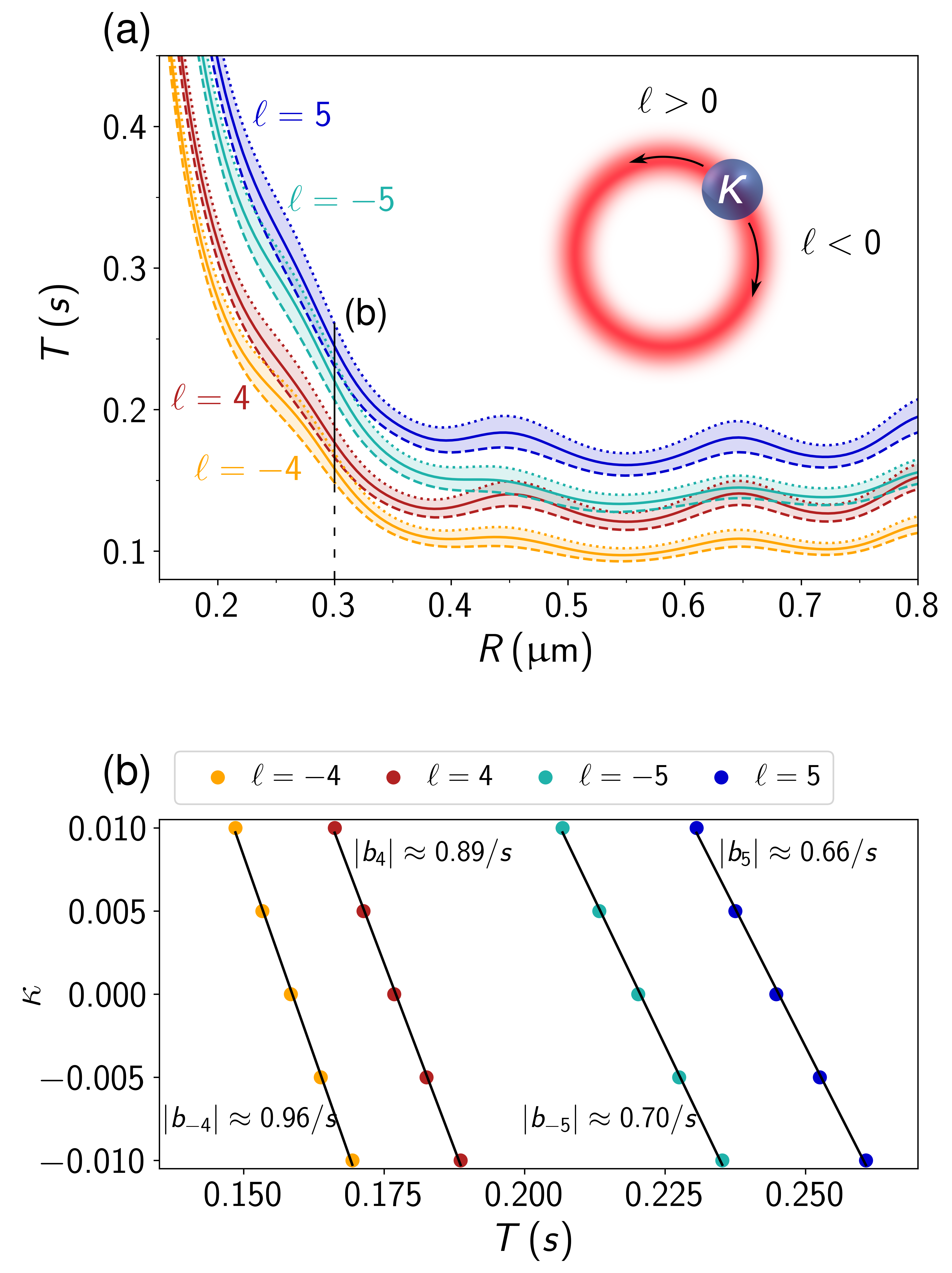}
    \caption{(a) Period $T$ as a function of the radius $R$ of a microsphere trapped by a vortex beam with topological charge $\ell=-4$ (orange), $\ell=4$ (red), $\ell=-5$ (cyan) and $\ell=5$ (blue). 
    As illustrated by the inset, the rotation direction is defined by the sign of the topological charge.
    We calculated the periods for spheres with different chirality index $\kappa$.
    The solid curves correspond to the case of an achiral sphere ($\kappa =0$), while the dashed and dotted curves correspond to chirality indices of $0.01$ and $-0.01$, respectively.
    The shaded area, bounded by the dotted and dashed lines for each topological charge, contains the period of spheres with $\kappa$-values between the two limiting cases. 
    In the considered interval, the period is linearly decreasing as a function of $\kappa$, as exemplified in (b) for $R=0.3\microns$ with the same $\ell$-values. 
    We performed a linear fit (black curves) with the absolute values of the slopes $b_\ell$ displayed at each curve.
    }
    \label{fig:T_x_r}
\end{figure}

Fig.~\ref{fig:T_x_r}(a) depicts the period of chiral and non-chiral spheres as a function of their radius for Laguerre-Gaussian beams of topological charges $\ell = \pm4$  and $\ell = \pm5$. These values were chosen so that the period could be shown for a variety of sphere radii. If the topological charge is too small, larger particles will be trapped on the beam axis. \cite{Fonseca2023}
For all cases, the period for spheres with chirality index $\kappa = 0.01$ (dashed lines) and $\kappa = -0.01$ (dotted lines) is shown, as well as the period for a non-chiral sphere (solid lines). 
The shaded area between the dotted and dashed curves accounts for the period of spheres with a chirality index in the interval $-0.01< \kappa < 0.01$.
Notice that the curves for topological charges with the same absolute value but different signs are not the same. 
This happens because we are considering a left-circularly polarized beam before focusing ($\sigma = 1$), thus breaking the symmetry between the $\pm \ell$ cases even for achiral spheres. 
Indeed, when the topological charge is positive, the orbital angular momentum has the same sign as the spin angular momentum, while for negative topological charges, the sign is opposite. 

Independently of the topological charge, all the curves exhibit the same general behavior. 
For radii $R$ smaller than about $0.35 \upmu$m, the period monotonically increases as the radius decreases, while for larger radii, it exhibits oscillations. 
This can be understood in terms of a decomposition of the optical force into a conservative and a non-conservative component. 
The conservative component, usually called the gradient force, pulls the particle towards the region of maximum intensity. 
On the other hand, the non-conservative component, usually called the scattering force, arises from radiation pressure and from the field's non-uniform helicity \cite{Albaladejo2009}.
Since the beam before focusing is circularly polarized, the azimuthal component $Q_{\phi}$
does not depend on $\phi$, by azimuthal symmetry. 
Thus, the line integral of the optical force along a closed circle around the optical axis is proportional to $Q_{\phi},$ 
showing that this component is non-conservative.
When the particle radius is small compared to the wavelength of the light ($R\ll \lambda$), 
its scattering can be well described in the Rayleigh limit. In this limit, the conservative component is proportional to the gradient of the electric energy density and dominates the non-conservative one, which explains the strong suppression of $Q_{\phi}$ and the resulting increase in the period as the radius decreases. 
%in which only the gradient component of the optical force is responsible for trapping, while the scattering component pushes the particle in the beam propagation direction. 
On the other hand, the non-conservative contribution builds up as the radius increases and becomes comparable to the wavelength ($R\approx \lambda_0/n_{\rm w}$) in the Mie scattering regime, giving rise to an azimuthal force component that drives the particle on its orbit. 
The oscillations shown in Fig.~\ref{fig:T_x_r}(a) are a consequence of interference effects inside the sphere.\cite{MaiaNeto2000}

In spite of the overall similar behavior discussed above, Fig.~\ref{fig:T_x_r}(a) shows a clear split between the curves corresponding to $\kappa=-0.01$ (dotted) and $\kappa=0.01$ (dashed).
The rotation period decreases monotonically with the chirality index as is exemplified in Figure~\ref{fig:T_x_r}(b) for a sphere of radius $R = 0.3 \microns$. 
An approximately linear relationship exists between the chirality parameter and the period for a fixed topological charge.
Variations in the chirality index $\delta \kappa_\ell$ are thus directly proportional to variations of the period $\delta T_\ell$, \textit{i.e.}
\begin{equation}\label{eq:delta_kp_T}
    \delta \kappa_{\ell} = |b_{\ell}| \delta T_\ell\, ,
\end{equation}
where $b_\ell$ is the slope of the linear fit of the $\kappa(T)$-curves as illustrated in Fig.~\ref{fig:T_x_r}(b).
Notice that we have added an index $\ell$ to the error in period measurements $\delta T_\ell$. Since the period monotonically increases with $|\ell|$ \cite{Curtis2003}, a fixed uncertainty would mean that the precision at higher topological charges is greater than that at smaller ones. Then, any gain in the resolution $\delta \kappa_{\ell}$ could be considered as an artifact of assuming a progressively smaller relative uncertainty. In order to allow for a fair comparison between different modes, we assume that the period is measured with the same  relative uncertainty $\xi$ for all modes and define
\begin{equation}\label{dT_relative}
    \delta T_{\ell} = \xi \overline{T_{\ell}}\, ,
\end{equation}
where $\overline{T_\ell}$ is the average period for the mode $\ell$ in the considered $\kappa$ interval from $-0.01$ to $0.01$.

\begin{figure}
    \centering
    \includegraphics[width=0.45\textwidth]{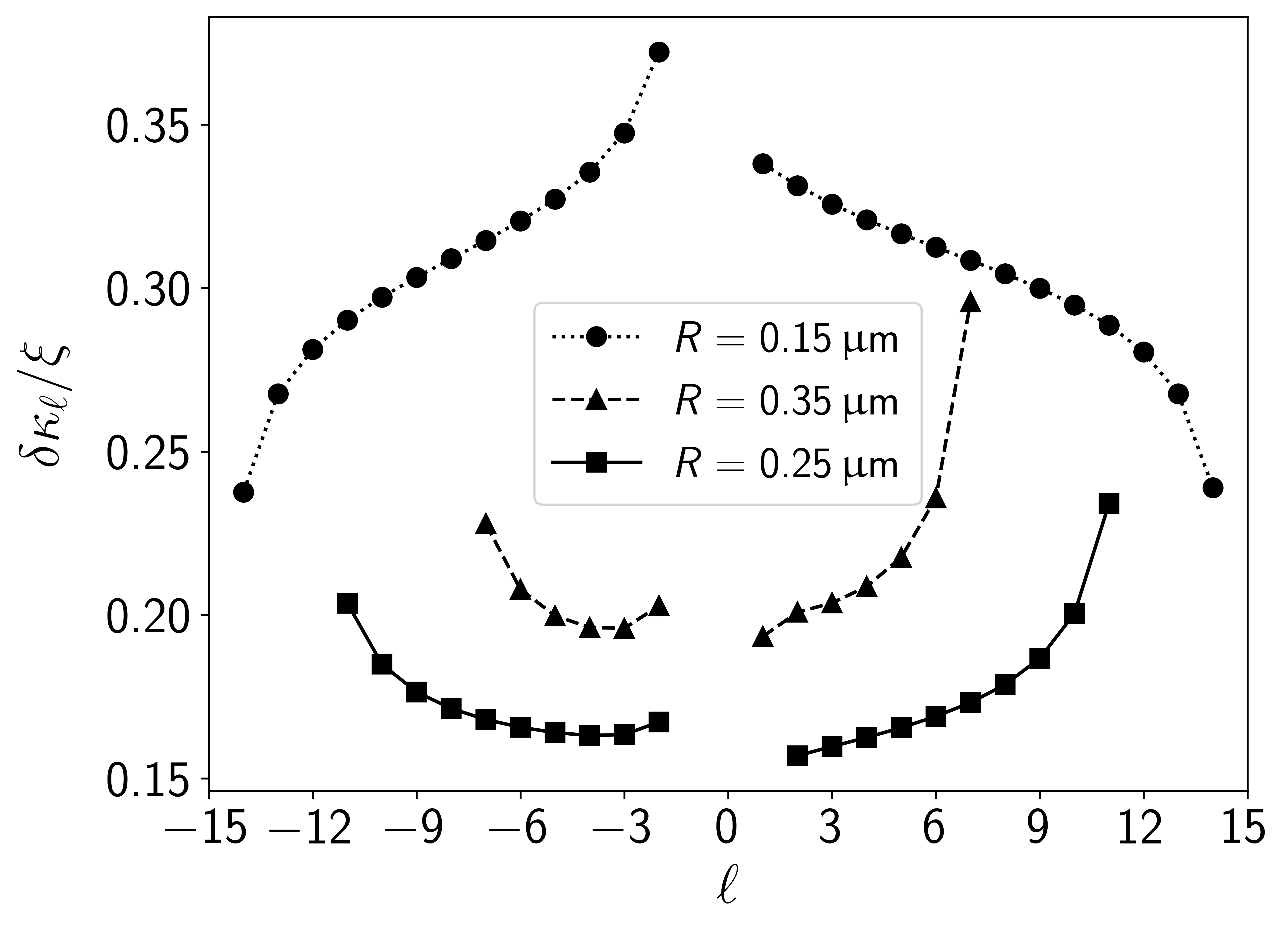}
    \caption{Minimum measurable chirality index $\delta \kappa_\ell$ scaled by the relative period uncertainty $\xi$ as a function of the topological charge for spheres with radii $0.15 \microns$ (circle symbols), $0.25 \microns $ (square symbols), $0.35 \microns$ (triangle symbols). 
    The connecting lines serve as visual guides.
    }
    \label{fig:dKp_x_Ell}
\end{figure}

Using the definition \eqref{eq:delta_kp_T}, we have investigated quantitatively the $\kappa$-resolution that could be achieved through period measurements.
Figure~\ref{fig:dKp_x_Ell} displays $\delta \kappa_\ell$ as a function of the topological charge $\ell$ for beads of radii $0.15$, $0.25$ and $0.35\microns$ scaled by the relative error of the period $\xi$. 
In each case, we plot all the values of $\ell$ for which a well-defined on-ring position exists.
For values of $\ell$ below those displayed in the set of points corresponding to each radius, the respective particle would be trapped on the optical axis. 
On the other hand, for values of $\ell$ greater than those presented, no point in space satisfies Eqs.~\eqref{Eq:EqZcondition} and \eqref{Eq:EqRhocondition} simultaneously for the given radius values, and no optical trapping is possible.
Notice that this upper limit for the available $\ell$ values can only exist in the Mie scattering regime, where the scattering component of the optical force plays an important role.
Indeed, in the Rayleigh regime the gradient force will necessarily pull the particle towards the ring of maximum intensity.
This is in accordance with the fact that the number of available topological charges decreases as the particle becomes larger.

For $R = 0.15 \microns$, the precision in $\kappa$ measurements monotonically increases with $|\ell|$. 
The lowest $\delta \kappa_\ell$ value achieved was $\delta\kappa_{-14} \approx 0.24 \xi$. 
On the other hand, the results for $R = 0.25 \microns$ and $R = 0.35 \microns$ show that $\delta \kappa_\ell$ finds its minima for much lower values of the topological charge. 
This can be advantageous, since the period for these modes is much smaller at the same power, allowing for good statistics with less acquisition time. 
The lowest values of $\delta \kappa_\ell$ obtained for $R = 0.25 \microns$ and $R = 0.35 \microns$ were $\delta\kappa_{2} \approx 0.16 \xi$ and $\delta\kappa_{1} \approx 0.19 \xi $, respectively. 
Hence, for a relative uncertainty $\xi=10^{-3}$ of the period measurement\cite{Chen2018}, we would find a precision of the order of $10^{-4}$ for the chirality measurement. 
Due to the linear relation \eqref{eq:delta_kp_T}, improving the precision of period measurements would lead to a proportional enhancement in the chiral resolution of our method.

\begin{figure}
    \centering
    \includegraphics[width=0.4\textwidth]{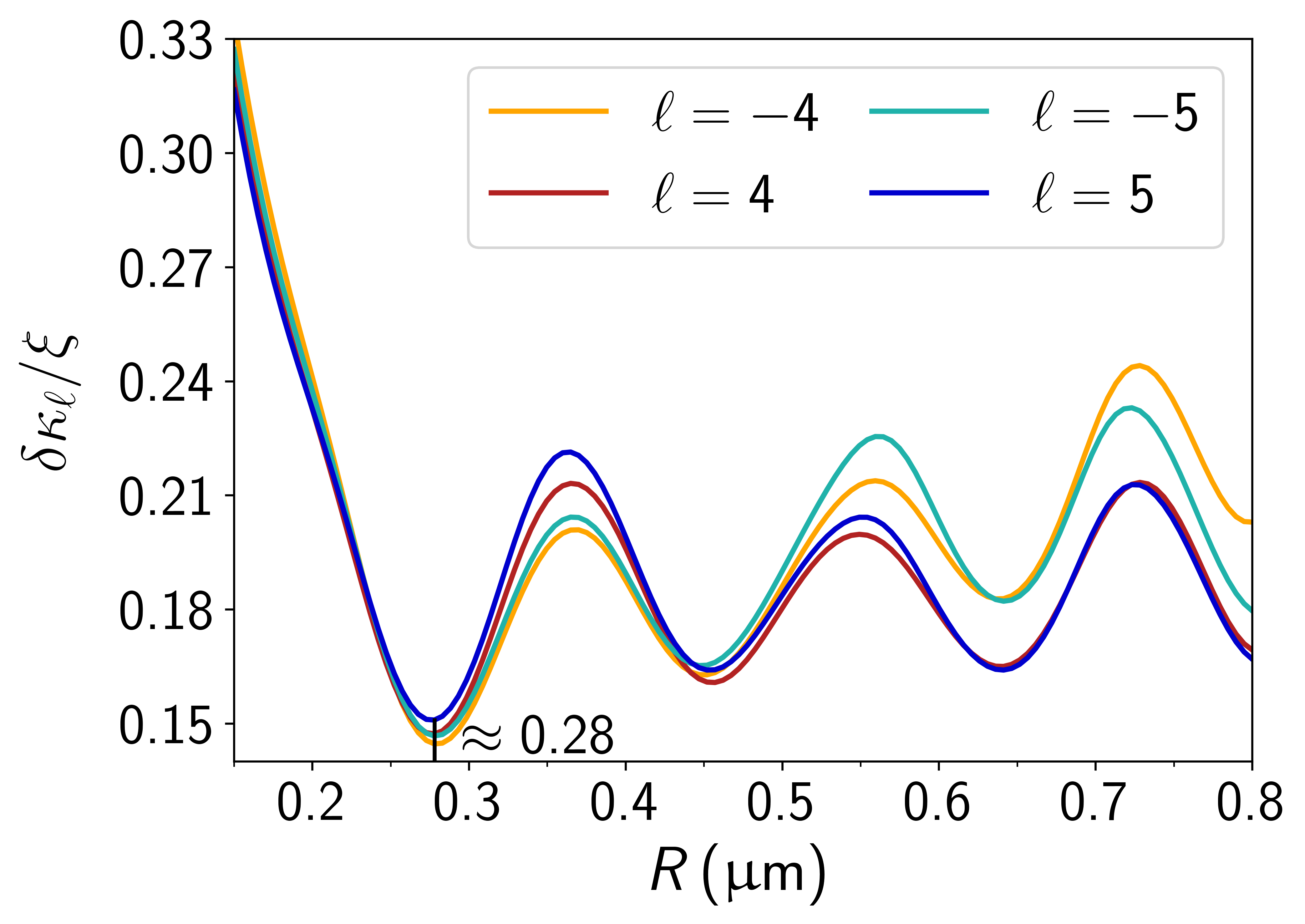}
    \caption{Resolution $\delta \kappa_\ell/\xi$ of the chirality index as a function of the sphere radius $R$ for the same topological charges as in Fig.~\ref{fig:T_x_r}. Global minima are identified near $R \approx 0.28\microns$.
    }
    \label{fig:dKp_x_R}
\end{figure}

It is worth noting that a change in radius appears to displace the points globally, \textit{i.~e.}, it either enhances or diminishes precision across all $\ell$ values for the examples shown in Fig.~\ref{fig:dKp_x_Ell}. 
Also, there is no monotonic relationship between $\delta \kappa_\ell$ and $R$. 
The general $\kappa$-resolution enhances from $R = 0.15 \microns$ to $R = 0.25 \microns$, but worsens from the latter to $R = 0.35 \microns$. 
This suggests the existence of an optimal radius for which the period is most sensitive to $\kappa$. 
In order to find such a value, we have performed a calculation of $\delta \kappa_\ell/\xi$ as a function of the particle radius for fixed topological charges $\ell = \pm 4$ and $\ell=\pm5$, and the results are depicted in Fig.~\ref{fig:dKp_x_R}. 
The curves exhibit global minima near $R \approx 0.28 \microns,$ which 
seems to be the radius allowing for the most sensitive chirality measurement. 
In the region $R < 0.28 \microns$, the resolution progressively worsens as the microsphere becomes smaller. 
On the other hand, in the region $R > 0.28 \microns$, $\delta \kappa_\ell$ exhibits an oscillatory behavior, meaning that the resolution of the measurement is of the same order of magnitude for particles within that region.

We would like to highlight an interesting aspect of Eq.~\eqref{eq:delta_kp_T}. 
Since the
period scales with power as $T\sim 1/P,$ 
the slope $b_{\ell} = \left(\partial \kappa/\partial T\right)_{\kappa = 0}$ goes as $b_{\ell} \sim P,$
and then $\delta \kappa_\ell/\xi$, as defined by
Eqs.~\eqref{eq:delta_kp_T} and \eqref{dT_relative},
is power-independent, and so are the arguments developed throughout this section. 
Hence, in a real implementation, the power can be chosen such that the precision of the period measurement is maximized.
For example, when using large values of $|\ell|$, one may freely increase the power in order to reduce the period and thus reduce the data acquisition time necessary to perform a good statistical analysis. 
In addition, increasing the power also reduces the effect of the particle's Brownian fluctuations, allowing for more precise determinations of periods, and thus providing a greater chiral resolution.

\begin{figure}
    \centering
    \includegraphics[width=0.45\textwidth]{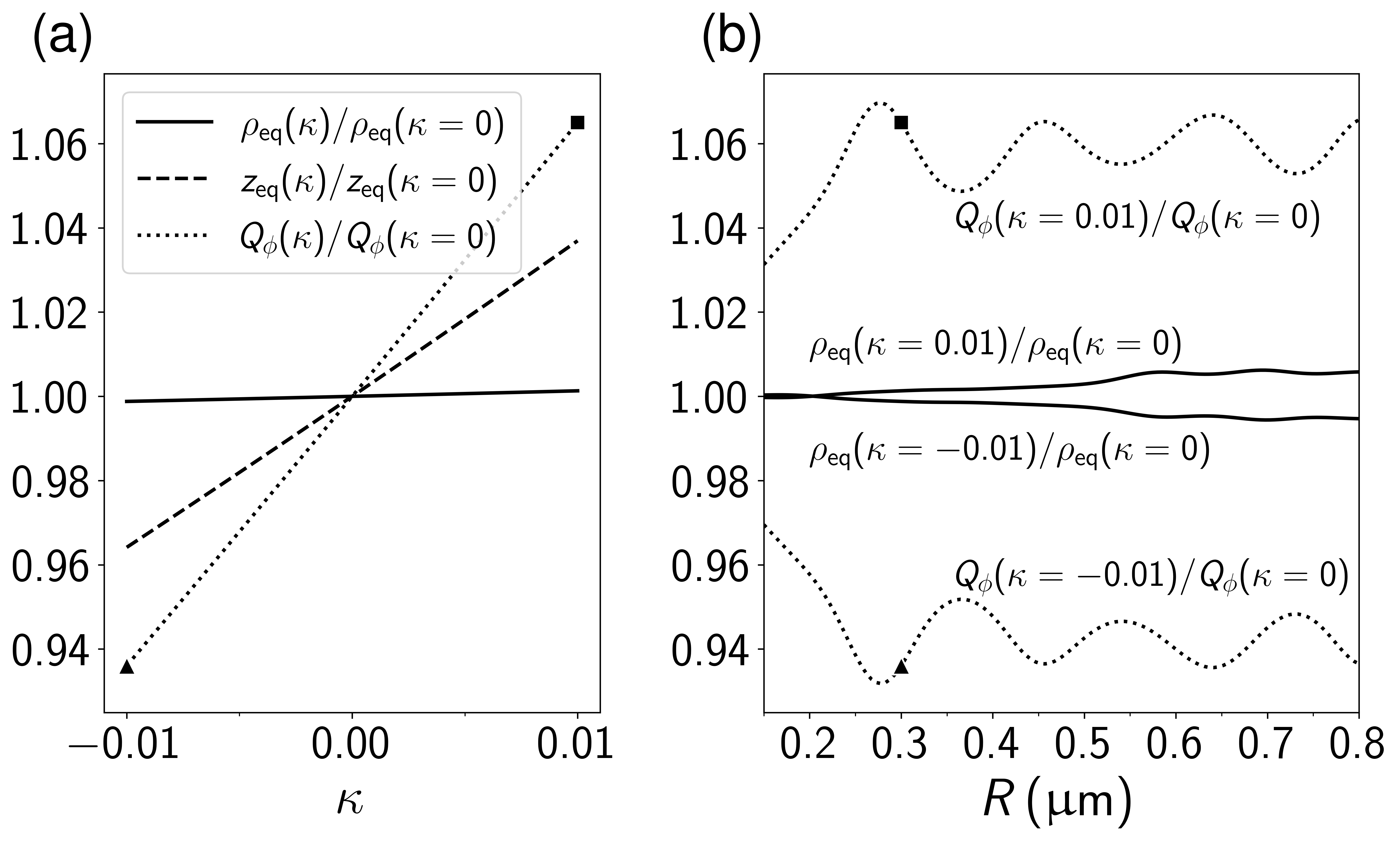}
    \caption{(a) Cylindrical coordinates of the stable orbit ($\rho_\equi$, $z_\equi$) and azimuthal force efficiency ($Q_\phi$) as functions of the chirality index $\kappa$. 
    The results are shown for a sphere of radius $R=0.3\microns$ and vortex beam with $\ell=4$. Each quantity is normalized by the respective value in the achiral case ($\kappa = 0$).
    (b) Radius of the orbit and azimuthal force efficiency for fixed chirality indices ($\kappa=\pm 0.01$) as functions of the sphere radii. As in (a), the results are shown for a vortex beam with $\ell=4$. The connection to the results for the azimuthal force in (a) is highlighted by the square ($\kappa = 0.01$) and triangle ($\kappa=-0.01$) symbol.}
    \label{fig:eq_x_kp}
\end{figure}

In Ref.~\citenum{Li_2021}, the authors show that, for particles confined to the focal plane, the average orbital radius depends on $\kappa$. 
Inspired by their work, we have also investigated the possibility of characterizing a particle's chirality through the orbital radius, rather than using the period.
In Fig.~\ref{fig:eq_x_kp}(a), the orbit coordinates $\rho_\equi$ and $z_\equi$ as well as the azimuthal force efficiency in the ring regime $Q_\phi = Q_\phi(\rho_\equi, z_\equi)$ are shown as functions of $\kappa$, normalized by their value at $\kappa = 0$. In the represented interval, all quantities exhibit linear behavior, but it can be observed that the azimuthal force varies more rapidly than the orbital radius. This fact is not just a particularity of the chosen radius, as can be seen in Fig.~\ref{fig:eq_x_kp}(b), where we plot the same relative quantities as functions of the sphere radius for different chirality indices. From Eq.~\ref{Eq:period}, we see that the period is proportional to $\rho_\equi$ and inversely proportional to $Q_\phi$, and therefore the dependence of the period on $\kappa$ is mainly due to $Q_\phi$. Thus, a measure of $\kappa$ based solely on the orbital radius, even if done with the same precision as period measurements, would necessarily have lower chiral resolution than a measurement made through the period. The stronger variation of the azimuthal force with the chirality index also explains why the period shown in Fig.~\ref{fig:T_x_r} decreases with increasing $\kappa$.
Moreover, it should be noted that highly precise measurements are easier to perform for the period than for the radius of the orbit.
By measuring the back\cite{Schaffer2007} or forward-scattered~\cite{Polimeno2018} light and Fourier transforming the signal to produce a power spectrum, one can extract the orbital frequency.
In contrast with the radius, the period depends on externally tunable parameters like the beam power, the Stokes drag coefficient, and the topological charge, thus allowing for optimizing the measurement.

\section*{Conclusions}

In conclusion, we have introduced a method to measure the chirality index of micro-sized particles with a precision up to $10^{-4}$. 
The method is based on measuring the period of a particle trapped by a focused vortex beam. 
The resolution does not depend on the power of the beam.
Compared to similar existing proposals \cite{ali2020sphere} our method offers a gain of at least one order of magnitude in precision. This result is of particular interest for probing and characterizing the chiroptical response of single, individual particles that typically exhibit weak light-chiral matter interactions. The precision of this method can be further improved if period measurements with a relative error below $10^{-3}$ can be achieved. Our findings may have applications in enantioselection of particles with very small chiral indexes, such as particles made of naturally occurring materials. 

%\section*{Author Contributions}...

\section*{Conflicts of interest}
The authors declare no conflicts of interest.

\section*{Acknowledgements}

We are grateful to Cyriaque Genet, Guilherme Moura, Leonardo Menezes, and Paula Monteiro for fruitful discussions. 
P.~A.~M.~N.,  N.~B.~V., and F.~A.~P. acknowledge funding from the Brazilian agencies Conselho Nacional de Desenvolvimento Cient\'{\i}fico e Tecnol\'ogico (CNPq--Brazil), Coordenaç\~ao de Aperfeiçamento de Pessoal de N\'{\i}vel Superior (CAPES--Brazil),  Instituto Nacional de Ci\^encia e Tecnologia de Fluidos Complexos  (INCT-FCx), and the Research Foundations of the States of Rio de Janeiro (FAPERJ) and S\~ao Paulo (FAPESP).

%%%REFERENCES%%%
\bibliography{rsc}
\bibliographystyle{rsc} 

\end{document}